\documentclass[11pt,twoside]{article}
\usepackage{asp2014}

\begin{document}

\title{
The ISM at high redshifts: ALMA results and a look to the future}
\author{Andrew W. Blain$^1$ 
\affil{$^1$
University of Leicester, Physics \& Astronomy, University Road, Leicester, LE1 7RH, UK; 
\email{ab520@le.ac.uk}}}

\paperauthor{Andrew~W.~Blain}{ab520@le.ac.uk}{ORCID_Or_Blank}{University of Leicester}{Physics \& Astronomy}{University Road}{Leicester}{LE1 7RH}{UK}

\begin{abstract}
ALMA is revolutionizing the way we study and understand the astrophysics of galaxies, both as a whole and individually. By exploiting its unique sensitivity and resolution to make spatially and spectrally resolved images of the gas and dust in the interstellar medium (ISM), ALMA can reveal new information about the relationship between stars and gas, during and between galaxies' cycles of star formation and AGN fueling. However, this can only be done for a modest number of targets, and thus works in the context of large samples drawn from other surveys, while providing parallel deep imaging in small fields around. Recent  ALMA highlights are reviewed, and some areas where ALMA will potentially make great contributions in future are discussed. 
\end{abstract}

\section{Introduction} 

At the Dec 2014 Tokyo meeting, Edo Ibar presented the results of deep ALMA continuum surveys, Nick Scoville discussed ALMA's role in measuring dust in galaxies, and Linda Tacconi was expected to discuss the progress and future role of ALMA in understanding gas in the interstellar medium (ISM) of galaxies, since Linda could not attend, I presented on this area, and 
am contributing this part of the proceedings.

The ISM has always been a factor in observational astronomy. Absorption is clear to the naked eye in the Milky Way, and it provides some of the most visually striking features of nearby galaxies. Spectroscopic signatures of diffuse material in emission and absorption have been seen whenever new spectroscopic instruments have been fielded, and in almost every new class of galaxy found. The importance of understanding the effects and details of the ISM has been highlighted in the past year, as subtle features of polarized dust emission have very-publicly confounded early claims of the detection of an unambiguous signal of cosmic inflation. 

ALMA is the first telescope with the sheer size to be able to image the ISM quickly in almost any type of high-redshift galaxy. Previous interferometers could generally detect the most luminous high-redshift galaxies, but in observing campaigns that would span several days of integration time, from the earliest (e.g. Solomon et al.\ 1992; Frayer\ et al.\ 1998, 1999; Downes\ et al.\ 1999), to the most recent (e.g. Walter\ et al.\ 2012; Huynh\ et al.\ 2014). Before ALMA, the detections have all been at a relatively modest significance, and while they provided important positional information, redshifts, total luminosities and a view of galaxy-wide ISM properties, they have not typically revealed very detailed astrophysical information about the galaxies involved. 

The technology available for receivers and correlators has advanced tremendously, in part to deliver the ambitious requirements of ALMA. The first interferometers sensitive enough to detect distant galaxies' ISM could do so over bandwidths of only a fraction of a GHz, and were thus barely able to span the whole width of a molecular or atomic line from a galaxy's ISM. Advances made leading to ALMA have boosted the performance of existing interferometers, for example enabling the NOEMA project to expand the IRAM PdB interferometer, whose technical specifications exceed those of ALMA at present. Tremendous increases in bandwidth have provided spectrographs that have made existing mm-wave facilities more competitive: for example, the EMIR wideband multi-band spectrograph at the IRAM 30M telescope has provided a great deal of insight into the ISM of nearby galaxies (Saintonge et al. 2011), and helped to expand the line spectral energy distributions (LSED) for a range of galaxies (e.g.\ Wei\ss\ et al.\ 2005; Carilli \& Walter\ 2013). 

\section{ALMA and the ISM in distant galaxies}

There are multiple components to the ISM, and it is dynamic, with the production of energy in stars and from AGN affecting the physical state and motions of gas within and around the host galaxy. These energy sources are themselves controlled by the supply and processing of ISM gas. In relatively quiescent systems, and in undisturbed parts of interacting and active galaxies, dense molecular gas should trace the potential of the galaxy/halo. However, feedback from active regions can drive winds, at speeds exceeding 1000\,km/s from AGNs, to escape the galaxy halo. Mechanical energy from radio jets can also impact the ISM, with consequences for fueling AGN and star-formation, while the ongoing rain of gas from the intergalactic medium into the galaxy, and right down onto an AGN might potentially be visible to molecular line observations (e.g. Combes et al.\ 2013, 2014). 

Understanding the state of the ISM requires at least several CO lines to be observed to build up an LSED (e.g. Wei\ss\ et al. 2005). Probing high-density tracers (e.g. Imanishi \& Nakanishi 2014; Viti et al.\ 2014) can better resolve the distribution of gas within the galaxy. Furthermore, optical depth effects mean that observations of isotopes of CO might be required to sample the ISM in full. Hence, deep observations with multiple tunings are required, to compile broadband (sub)mm-wave spectra. Atmospheric transmission windows also make it impossible to observe all galaxies over the same range of wavelengths, and so multi-line studies are constrained to specific redshift ranges; see
the recent review by Carilli \& Walter\ (2013). 

The most complete LSED of the ISM in distant galaxies comes from a composite spectra obtained from bright gravitationally-lensed galaxies (e.g Spilker et al. 2014). While there are effects to consider about the uniformity of these samples, and how differential magnification affects the SED/LSED (Blain 1999; Serjeant 2012), the picture of a rich range of lines being probed is clear. 

The ISM of distant galaxies has also been probed by the {\it Spitzer} and {\it Herschel} space missions, and at longer wavelengths from the ground using SPT, ACT, LMT, the IRAM 30M, JCMT, APEX, CSO and ASTE. There are now wide fields with spectral coverage providing accurate SEDs for distant dusty galaxies, the wavelenght range spanning from  starlight to longwards of the peak of the thermal dust emission in the far-infrared(IR). The total luminosity radiated by the ISM, powered by absorbed optical and ultraviolet radiation is now well known across a substantial part of the Universe's history (e.g. Casey et al.\ 2014). Large surveys comparing {\it Spitzer} stellar masses, optical spectroscopic properties, and total luminosities from {\it Herschel} have revealed a perhaps surprising statistical uniformity of galaxies' properties at $z \sim 1$-2 - the "Main Sequence (MS)" (Elbaz et al. 2011; Magdis et al.\ 2012; Saintonge et al.\ 2012; Tacconi et al.\ 2013). The way in which this is established, regulated and maintained is still not known in detail. Determining the key details of the gas fueling star formation within carefully-selected subsamples of MS galaxies with ALMA, in conjunction with resolved images and spectra in the near-IR, should reveal the processes in galaxies that lie on the relation. Furthermore, there is a comparable correlation in the range of metallicities of galaxies - the "Fundamental Metallicity Relation (FMR)" - that applies to bulk galaxy properties, and is inevitably important for understanding the net result of the processes at work in typical galaxies over the bulk of cosmic time (Mannucci et al.\ 2010; Maier et al.\ 2014); ALMA can observe typical distant galaxies on these relations, and the results can be compared with existing large-scale surveys of the cold gas component of nearby galaxies (e.g. Santoinge et al. 2011).

Programs are underway to image galaxies out to large cosmic distances over contiguous areas that are large enough to overwhelm ALMA's mapping speed. The area of sky with excellent imaging quality in depth, bands and resolution, ranges from the CANDELS coverage of exquisite angular resolution over a fraction of a square degree using {\it HST} to the many square degrees of sky that will soon be covered to the maximum depth that can be reached by the 1.5-deg$^2$-field Subaru Telescope's Hyper Suprime-Cam instrument. Euclid and WFIRST plan very wide high-resolution surveys. There are degree-sized deep surveys by ground-based (sub)mm telescopes IRAM 30M, ASTE and JCMT, and 1000-deg$^2$ fields surveyed by {\it Spitzer} and {\it Herschel}. The SPT and ACT ground-based CMB telescopes probe thousands of square degrees of the mm sky, and over the whole sky there are mid-IR and mm-wave surveys from WISE and Planck Surveyor. New wide/deep radio imaging capabilities are arriving from SKA precursors. 

\section{The unique capabilities of ALMA} 

ALMA's excellent spatial resolution allows high-redshift galaxies to be investigated, even though their ISM continuum emission is too faint to be detected above confusion noise using lower-resolution telescopes. While spectral-line observations reduce confusion, serendipitous discoveries of line-emitting galaxies are rare, and no single-dish telescopes can resolve individual high-redshift galaxies. 

ALMA can reveal serendiptious redshifts for galaxies whose location might not be known sufficiently accurately for conventional optical/near-IR spectroscopy. A prototype is the case of the brightest submillimeter galaxy (SMG) found in the Hubble Deep Field (Hughes et al. 1998), whose identification was unclear. An early IRAM PdBI image found an accurate position, but no 
convincing counterpart (Downes et al. 1999). A redshift was finally determined using the updated PdBI by Walter et al.\ (2012). 

One-to-one matching and submm detection of substantial numbers of galaxies identified in deep multiwaveband surveys, like significant stellar-mass-selected samples from CANDELS, demands ALMA observations, since only a handful of the galaxies within these fields are sufficiently bright to be probed using previous interferometers (e.g. Walter et al. 2014) in a reasonable time. The resolution is limited to match the requirement to obtain significant detections in each synthesized beam. ALMA's spectral coverage in all the submm windows from 3-0.3\,mm, resolving power, and sensitivity puts it in a strong position to resolve typical high-redshift galaxies, even the faintest objects in CANDELS, given enough time. 
ALMA could resolve the bulk of the ISM gas in a high-redshift galaxy, whether associated with a cool ordered disk, with turbulent active regions, with infall, or in the process expulsion by winds. It will reveal the location, relative excitation and temperature of different components of the ISM, provide dynamical evidence to distinguish merging subunits from systematic halo-wide rotation, and provide an accurate measurement of the dynamical mass under testable assumptions about whether the gas is virialized, has enhanced motions from turbulence or is supported by rotation. Observations of redshifted CO(1-0) from the Jansky VLA (JVLA) can be combined to provide a consistent high-resolution picture of the coolest molecular gas (e.g. Wagg et al. 2014). 

At present, this sort of investigation has been demonstrated for galaxies that have been made brighter and physically larger on the sky by gravitational lensing, most dramatically the "Eyelash" (Swinbank et al. 2010); see also Bothwell et al. (2013) and Spilker et al. (2014). ALMA's power will allow similar studies in galaxies that are not boosted by gravitational lensing, while large samples of bright lensed galaxies can be despatched very quickly using ALMA (Vieira et al. 2013; Messias et al. 2014), or even with smaller interferometers (Bussmann et al. 2013; Rawle et al. 2014).

The physical conditions in even the most heavily obscured regions can be probed using ALMA alone, by measuring molecular and atomic fine-structure line ratios (e.g. Meijerink et al.\ 2011; Tomassetti et al. 2014), and by observing rotational--vibrational and high-rotational-level molecular emission that can only occur in energetic regions (Imanishi \& Nakanishi 2013). By comparing the ALMA images of galaxies with those obtained from ground-based imaging near-IR spectrographs, {\it HST} in the optical, and radio interferometers, it should be possible to obtain an excellent picture of the motion of molecular, atomic and ionized gas in specific powerful galaxies, providing insight into the fuel available for, and both the duration and current efficiency of star formation and AGN fueling. The independent requirements of a nearby adaptive-optics tip-tilt star, lines accessible in ALMA atmospheric windows, and relative brightness means that gathering large samples of these galaxies would be challenging; however, studying individual examples remains important.

In deep images, ALMA can provide very high fidelity maps of the dynamics of gas in distant galaxies, probing molecular gas that is likely to be amongst the coldest component of the ISM, and thus to be the most accurate tracer of the gravitational field. It is possible that very precise measurements could perhaps probe the gravitational lensing shear field on small scales (Blain 2002a), or exploit negative lensing magnification in the cores of dense regions to increase the effective primary beam area for serendiptious surveys in cases where sensitivity is abundant (Blain 2002b). 

Any deep ALMA observation of a distant galaxy conducts an effective deep pencil-beam serendipitous spectroscopic survey throughout the primary beam. ALMA's survey performance in going deeper than previous facilities (in both line and continuum observations) has been highlighted by Hatsukade et al. (2013); also see Ibar (this work) and Carniani et al. (2015). ALMA is an almost ideal spectrograph for resolving the ISM in individual high-redshift galaxies and their immediate surroundings, especially those with redshifts that are already known, and in which line-of-sight velocity information is guaranteed. However, ALMA does not survey for potential companions on scales larger than the primary beam, and is thus not perfectly matched to compliment the arcminute-scale patrol radius of ambitious new optical/near-IR imaging spectrographs like MUSE and KMOS. On the other hand, ALMA is well matched to future ELT instruments, with excellent adaptive-optics correction over relatively small fields. 

\section{ALMA imaging results for infrared-bright and normal galaxies}

There are now huge numbers of galaxies selected by their ISM dust emission: the SMGs (Smail et al. 1997; Blain et al.\ 2002; Wei\ss\ et al. 2009; Karim et al. 2013), and their cousins selected at shorter wavelengths by Herschel, Spitzer and WISE: "distant dusty galaxies" (DDGs e.g. Casey et al.\ 2014). These galaxies are up to several hundred times more luminous than typical high-redshift galaxies, and are much rarer, by a factor of about 100 for the SMGs, and by up to 10$^4$ in the case of WISE-selected "HotDOGs" (Eisenhardt et al. 2012; Wu et al. 2014; Assef et al.\ 2015; Tsai et al.\ 2015). Nevertheless, these rare (and presumably short-lived) galaxies are bursting with an extreme rate of activity and must be associated with more mundane less-active hosts, that themselves probably lie on the MS and FMR relations.  ALMA can discriminate the ordinary and remarkable parts of these galaxies, and provide more complete size, dynamical, and excitation information than accessible at other wavelengths due to substantial extinction. ALMA's ability to resolve the structures, dynamics and physical conditions in the ISM should soon allow us to understand SMGs/DDGs in detail. 

Imaging high-redshift SMGs with ALMA (e.g. Karim et al. 2013) has revealed that an interesting number break up into multiple components. Perhaps this shows a limit to the surface brightness of galaxies, with all very luminous objects breaking up into interacting subunits, or perhaps many are even accidental superpositions. 

The most impressive haul of ALMA spectra so far has been for galaxies identified as gravitational lenses from wide-area multicolor mm-wave continuum surveys primarily aimed at measuring the abundance of Sunyaev--Zeldovich effect clusters (e.g. Hezaveh et al. 2013; Vieira et al. 2013; Wei\ss et al. 2013; Spilker et al.\ 2014); comparable samples are available from the Planck, Herschel and ACT telescopes. 

The first distant galaxies imaged by ALMA can be divided into several catagories: bright lensed objects (see above); SMGs and dusty AGNs at various resolutions (e.g. Swinbank et al.\ 2012; Nagao et al.\ 2012; Wagg et al.\ 2012; Carniani et al.\ 2013, Carilli et al.\ 2013; Hodge et al.\ 2013; Huynh et al.\ 2013; Karim et al.\ 2013; Wang et al.\ 2013b; De Breuck et al.\ 2014; Simpson et al.\ 2014, 2015; Thomson et al.\ 2014); more typical star-forming galaxies (e.g. Decarli et al.\ 2014; Ono et al.\ 2014; Ota et al.\ 2014; Riechers et al.\ 2014); high-redshift QSOs (e.g. Wang et al.\ 2013a; Gilli et al.\ 2014); and more unusual objects: gamma-ray burst (GRB) host galaxies (Berger et al.\ 2014; Hatsukade et al.\ 2014; Michalowski et al.\ 2014), Lyman-$\alpha$ emitters (Ouchi et al.\ 2012; Williams et al.\ 2014), and low-metallicity dwarves (Hunt et al.\ 2014). 

\section{High-speed molecular outflows} 

Molecular material ejected by powerful low-redshift galaxies has been detected from the ground (Iono et al.\ 2007; Nesvadba et al.\ 2008; Feruglio et al.\ 2010; Alatalo et al.\ 2011; Aalto et al.\ 2012; Cicone et al.\ 2014), and space (Fischer et al.\ 2010; Sturm et al.\ 2011), in the broad wings of molecular line emission and in P-Cygni profiles of OH lines respectively. 
Powerful AGN at high redshifts also show broad molecular line wings in sensitive, broad-band mm-wave spectral images of redshifted relatively-bright fine-structure lines (Cicone et al.\ 2015). 

The detection of these broad lines requires substantial S/N detections of the narrower, bound ISM emission in order to be reach the necessary sensitivity. While providing otherwise unavailable insight into the mechanical energy output by luminous galaxies, they can thus also detail the less-dramatic but comparably interesting dynamics of gas in the host galaxy and any nearby companions. The first such ALMA results are appearing (Garcia-Burillo et al. 2014; Sun et al. 2014). 

\section{Large-scale structure and probing galaxy mass scales}

There are various indications that, on a variety of spatial scales, luminous dusty galaxies are strongly clustered, whether from spectroscopy of counterparts (Blain et al. 2004), from correlation functions using {\it Herschel} surveys (Viero et al. 2013), from the overdensity of ultraluminous SMG companions in the fields of WISE-selected galaxies from ground-based images (Jones et al. 2014, 2015), or within the ALMA primary beam: both in follow-up images of SMG surveys (e.g. Karim et al. 2013), or around WISE--radio-selected AGNs (Lonsdale et al.\ 2015; Silva \& Sajina 2015; this volume). 

The modest field of view of ALMA is always likely to make it more powerful for the study of individual galaxies than for  conducting large surveys. However, because of its sensitivity and spatial resolution, and the absence of confusion noise, discovering serendipitous galaxies in ALMA fields will be a regular occurence (Tamura et al.\ 2014). 

ALMA is not ideally suited to studying galaxy clustering on scales larger than the primary beam; however, it does offer the opportunity to test whether the detailed astrophysics of galaxies depend on their environments, which is likely to be important for understanding the development of the MS and FMR relations, and potentially to reveal the astrophysics responsible for the downsizing process. It can also provide high-resolution extinction-free images for modelling gravitational lenses, and reveal structure otherwise too small to probe (e.g. Vieira et al.\ 2013). 

\section{Future capabilities?}

Owing to the modest 10-km/s resolution necessary to study galaxies, there are few demands from this science area on the spectral resoution of the ALMA correlator. However, outflows that are several 1000 km/s in width do challenge the available bandwidth coverage at the very highest frequencies. There are closely separated lines from different species that could be imaged simultaneously if the bandwidth could be increased to span a larger fraction of an atmospheric window, which would also increase the chances of success in serendipitous line (and continuum) surveys in parallel to deep observations. 

Absorption line studies against powerful continuum point sources would also potentially benefit from an increase in bandwidth (e.g.\ Wiklind \& Combes 1998). Measurements of very precise frequency offsets in absorbers can be used to limit changes in fundamental constants, and species that are far too rare to be detected in emission can be measured in absorption (Muller et al.\ 2007, 2014; Bagdonaite et al. 2012). The precision of ALMA spectroscopy and frequency accuracy also potentially allows some such tests using emission lines (e.g. Lentati et al. 2013). 


\section{Summary}
ALMA's first three years of operations, during which the time available on the array and its capabilities have been much less than will be ultimately delivered 
has shown its remarkable capabilities. 
ALMA will provide a vital resource for studying galaxies in the years ahead, providing 
data of exceptional and quite unprecedented quality. 


\acknowledgements I thank Linda Tacconi, Caitlin Casey, Ian Smail, Mark Swinbank and Fabian Walter for providing material for this presentation, and the organizers for such a wide-ranging program of highlights in such an excellent venue. AWB is supported by a Royal Society Wolfson Research Merit award.

\end{document}